\documentclass{vgtc}                          

\ifpdf
  \pdfoutput=1\relax                   
  \usepackage{graphicx}                
  \DeclareGraphicsExtensions{.pdf,.png,.jpg,.jpeg} 
\else
  \usepackage{graphicx}                
  \DeclareGraphicsExtensions{.eps}     
\fi%

\graphicspath{{figures/}{pictures/}{images/}{./}} 

\usepackage{microtype}                 
\PassOptionsToPackage{warn}{textcomp}  
\usepackage{textcomp}                  
\usepackage{mathptmx}                  
\usepackage{times}                     
\usepackage{cite}                      
\usepackage{tabu}                      
\usepackage{booktabs}                  
\usepackage{enumitem}
\usepackage{float}
\usepackage{scalerel}
\PassOptionsToPackage{bookmarks=false}{hyperref}
\onlineid{0}

\vgtccategory{Research}

\vgtcinsertpkg

\title{iConViz: Interactive Visual Exploration of the Default Contagion Risk of Networked-Guarantee Loans
}

\author{Zhibin Niu$^{*}$, Runlin Li$^{*}$, Junqi Wu\thanks{e-mail:\{zniu, runlinli, wujunqi, jwzhang\}@tju.edu.cn, Jiawan Zhang is the corresponding author.},
Dawei Cheng\thanks{e-mail: dawei.cheng@sjtu. edu.cn},
Jiawan Zhang$^{*}$
\vspace{-10pt}
}

\affiliation{%
  \scriptsize
  $^*$College of Intelligence and Computing, Tianjin University\\
  $^{\dag}$Department of Computer Science and Engineering, Shanghai Jiao Tong University
}

\teaser{
  \vspace{-5pt}
 \centering
 \includegraphics[width=1\linewidth]{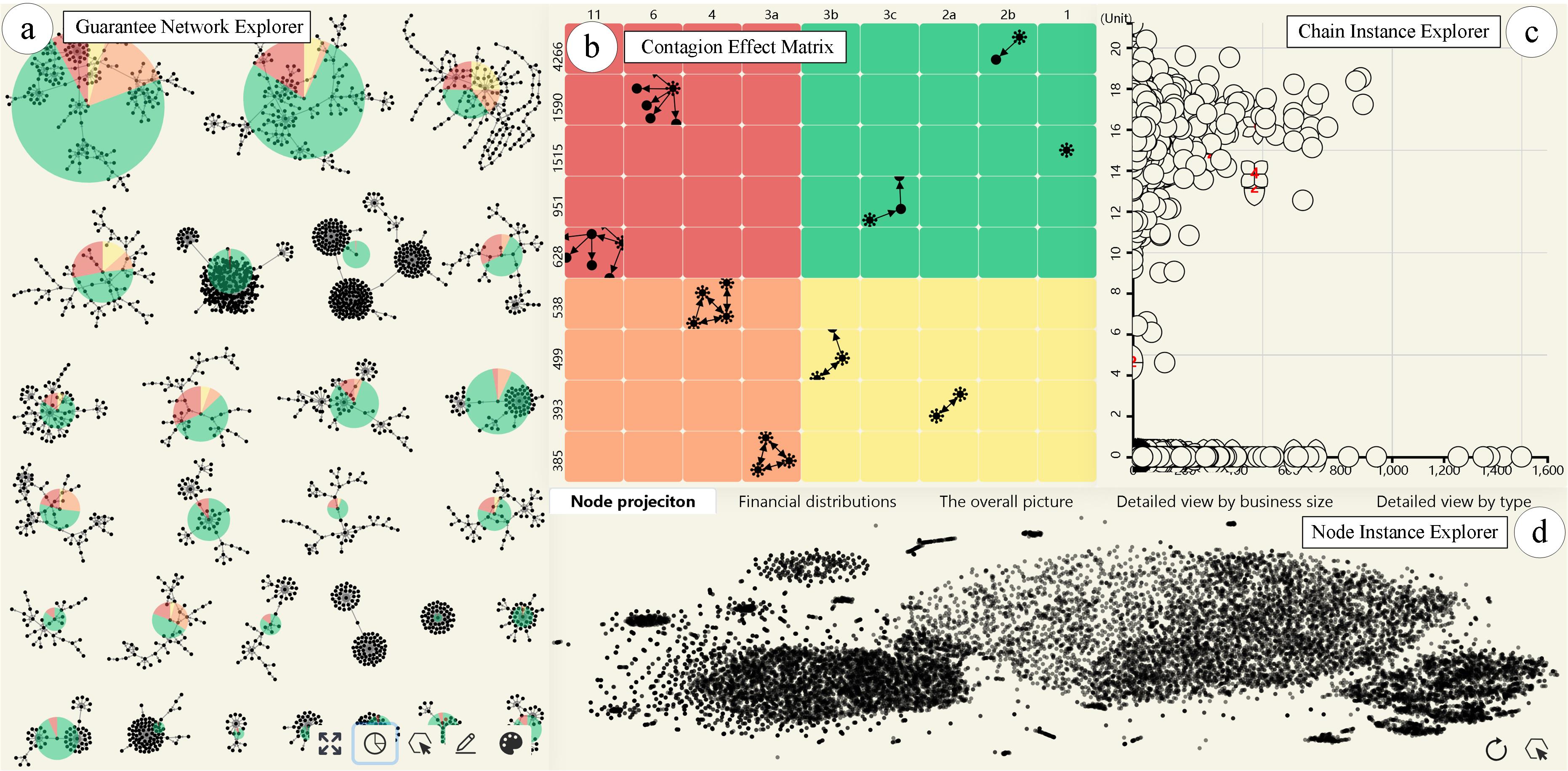}
 \vspace{-15pt}
  \caption{System interface of iConViz: (a) Guarantee Network Explorer. It facilitates an overview of and zooming in on levels of detail regarding guarantee networks using a network tessellation layout. It also offers intuitive metaphorical symbols (Contagion Effect Badges) of contagion risk to support the selection of interesting networks. (b) Contagion Effect Matrix. This gives detailed contagion risk patterns and quantifies severity. (c) Chain Instance Explorer. This supports further narrowing of the search space for instances of contagion from a financial perspective. (d) Node Instance Explorer. This visualizes the finest-grain information on demand. }
  \label{userinterface}
  }

\abstract{Groups of enterprises can serve as guarantees for one another and form complex networks when obtaining loans from commercial banks. During economic slowdowns, corporate default may spread like a virus and lead to large-scale defaults or even systemic financial crises. To help financial regulatory authorities and banks manage the risk associated with networked loans, we identified the default contagion risk, a pivotal issue in developing preventive measures, and established iConViz, an interactive visual analysis tool that facilitates the closed-loop analysis process. A novel financial metric, the contagion effect, was formulated to quantify the infectious consequences of guarantee chains in this type of network. Based on this metric, we designed and implemented a series of novel and coordinated views that address the analysis of financial problems. Experts evaluated the system using real-world financial data. The proposed approach grants practitioners the ability to avoid previous ad hoc analysis methodologies and extend coverage of the conventional Capital Accord to the banking industry.
} 


\begin{document}


\firstsection{Introduction}

\maketitle
Regulatory technology (RegTech) is designed to enhance transparency and consistency and address the regulatory challenges faced by financial services providers, including monitoring, reporting, and compliance obligations~\cite{arner2016fintech, arner2017fintech}. 
The rapid development of RegTech has increased awareness of visual analytics and artificial intelligence in this area~\cite{niu2020regvis}. The chief economist for the Bank of England, Andy Haldane, imagined the future of RegTech to be a global map of financial flow that charts spillovers and correlations~\cite{haldane2014managing}.

Networked-guarantee loans, a unique financial and banking phenomenon in some countries, are attracting increased attention from regulators and banks. When enterprises back one another in loan applications, they form complex directed networks. Highlighted by the multifaceted background of the growth period, the structural adjustment of the pain period, and early stages of the stimulus period, structural and deep-level contradictions have emerged in the economic development system. During economic slowdowns, corporate default  \footnote{Financial terminology referring to a business failure to fulfill an obligation, especially to repay a loan or appear in a court of law.} may spread like a virus and lead to large-scale defaults or even systemic financial crises. Thus, the need for risk management is more urgent than ever before. Monitoring the world’s financial status is so complicated that it is usually only after a capital chain rupture that regulators are able to study cases in depth. Regulators and banks are seeking to utilize data-driven and visual analytics approaches to managing the risk brought about by networked loans. However, the public research output for this problem is still relatively limited.

We worked closely with financial experts to conduct research on the guarantee network loan risk management problem. We identified that the default contagion risk for networked loans is an important yet unexplored interdisciplinary research problem. Data-driven and visual analytics-based approaches can provide fresh insights into assessing contagion risk and possible preventative measures. In this research, we propose iConViz, a novel visual analytics approach, as a means of helping financial experts conduct in-depth analyses of the default contagion risk problem. We believe that this is the first study to identify and formalize the contagion chain risk management problem for networked loans.

The main contributions of this work are as follows:
\begin{itemize}[itemsep= -4 pt, topsep = 0 pt]
\item  \emph{We discover and highlight eight interpretable contagion chain patterns} by analyzing real-world bank loan records. The patterns illustrate different contagion characteristics that lay the basis for quantitative contagion risk assessments (i.e., the contagion effect) and our visualization design.

 \item We propose a \emph{systematic data-driven and visual analytics approach} to the contagion risk problem within the framework of the closed-loop analysis process. Our approach offers financial experts the ability to avoid previously ad-hoc methodologies.



  \item We describe iConViz, an interactive visual analytics tool we developed to analyze the contagion risk problem with networked loans. Several novel visualization and interaction designs are proposed, such as the guarantee network tessellation layout, flower petal zooming interaction design that overcomes visual clutter, and the contagion effect badge that provides visual symbols for quantifying contagion risk.


\end{itemize}




\section{Related Work}

Data and visual analytics technologies are applied extensively in financial risk management problems in domains such as macro-prudential oversight and fraud detection in online transactions and investment~\cite{jeong2008evaluating,dumas2014financevis,FLOOD2016180,sarlin2016macroprudential}.  In this section, we first introduce the capital accord, an important risk management framework for the banking industry, and then works on interdisciplinary financial risk management and data visualization analytics.

\textbf{Capital Accord} The Basel Committee on Banking Supervision issued a series of recommendations on banking laws and regulations (Basel I, II, and III) to enhance the understanding of key supervisory issues and improving the quality of banking supervision~\cite{comitato2004international}. These principles have been widely accepted by banks around the world. Under Basel II, a series of parameters, also named as internal ratings-based approach~\cite{thomas2005interpreting}, are used to calculate the economic or regulatory capital of banking institutions. 1) Probability of default (PD). Default probability can be estimated from historical default data, observable prices of credit default swaps, bonds, and options on the common stock market identified using machine learning algorithms like decision trees, logistic regression, support vector machines, neural networks, genetic programming, ensemble methods, and many other machine learning processes~\cite{baesens2003benchmarking, louzada2016classification}. In credit risk management, the standard assumption is that a loan is considered in default when the client is past due on payment by at least three months. 2) Loss-given default (LGD). This refers to the share of an asset that is lost if a borrower defaults. LGD is facility-specific because such losses may be influenced by key transaction characteristics such as the presence of collateral and the degree of subordination. 3) Exposure at default (EAD). This is defined as the gross exposure of a facility upon the default of an obligor. 4) Expected loss. This can be formulated as the product of PD, LGD, and EAD.

\textbf{Regulatory visualisation} The rapid development of regulatory technology has raised awareness of information visualization and visual analytics in this area. We provided regvis.net, a visual bibliography of regulatory visualization for better indexing the literature~\cite{niu2020regvis}. In detail, many studies of visualization for macro-prudential oversight (measures systemic risk both timely and accurately) have been conducted for the financial sector~\cite{sarlin2016macroprudential}. The foremost work is by the volatility laboratory of Nobel Laureate Robert Engle. His group provides real-time online measurement, modelling, and forecasting of financial volatility and correlations with systemic risk via classic interactive charts such as bar charts, Box plots, map charts, and other prime examples~\cite{vlab}. In the cross-sector of machine learning and financial stability, Peter Sarlin published a series of works on visualized systemic risk analytics in cooperation with the European Central Bank, International Monetary Fund, and other international banks~\cite{visklab}. The self-organizing map, a neural network-based unsupervised learning method and visualization tool, can be utilized to evaluate and visualize financial stability with the power of simultaneous clustering and projection~\cite{sarlin2011visual,sarlin2013mapping,sarlin2013self}. It is also extended with a novel time dimension to decompose and identify temporal structural changes in macro-financial data around the global during financial crises ~\cite{sarlin2016macroprudential}. Fraud detection in transactions is always a primary concern of banks~\cite{dumas2014financevis}. The Advanced Detection System was one of the earliest risk visualization systems that monitored trades and quotations on the Nasdaq stock index, identifying patterns and practices of behavior of potential regulatory interest~\cite{kirkland1999nasd,leite2018eva}. Wirevis employs specific coordinated keyword visualization for wire transactions that can be used to detect suspicious accounts, transactions, behaviors~\cite{chang2007wirevis}. 3D tree maps have been introduced to monitor real-time stock market and identify unusual trading patterns, suspected traders (i.e., attackers), and attack plans~\cite{huang2009visualization}.
\vspace{-0.8pt}

Network visualization is employed to represent bank interrelations through financial discussion data~\cite{constantin2018network}. The force-directed layout is perhaps the most extensively utilized in the financial area. And the network centrality measurements such as node degree, betweenness, and closeness, K-core shell are used to measure and visualize the node or edge importance. For example, Rönnqvist and Sarlin collected text data from online financial forums and generated and visualized a co-mentioned bank network (i.e., interbank network), with which to quantify the bank interdependence (using centrality measurements), such as interbank lending and co-movement in market data~\cite{ronnqvist2015bank}. Xu and Chen discussed criminal network analysis and visualization, a very close domain. Their insights included that social network analysis can be used to analyze interaction patterns and criminal networks can be partitioned into subgroups of individuals by the centralities in their network~\cite{Xu,didimo2011an}. Bottom-up and top-down interaction are demonstrated can be effectively to reveal financial crimes such as money laundering and fraud in the financial activity network~\cite{didimo2011advanced}. Heijmans and others used animation to visualize and analyze the large transaction networks in the daily Dutch overnight money market~\cite{Heijmans2014Dynamic}. There were some other publications mining the subgraph structures and patterns to interpret the financial meaning. Among them, BitExTract was developed to observe the evolution of transaction and connection patterns of Bitcoin exchanges from different perspectives~\cite{yue2018bitextract}. A ego-centered node-link view depicts the trading network of exchanges and their temporal transaction distribution and facilitates the recognition of unique patterns.
\vspace{-0.8pt}

Guarantee networks consisting of multiple enterprises related by secured loans were first addressed by the computing community in 2015~\cite{meng2017netrating}. Subsequently, a risk management framework for these guarantee networks was introduced and a visual analytics approach presented~\cite{niu2018visual}. Since then, intensive research on default risk prediction has been performed~\cite{cheng2019dynamic,cheng2019risk}. However, to the best of our knowledge, the contagion risk management problem for guarantee network loans has not been adequately addressed. The present research is the first attempt to be made by financial computing research community. One core question is how to assess the risk of contagion introduced by a network structure. The metric contagion effect, together with a practical visual analysis pipeline, is thus proposed to facilitate a better understanding and more accurate assessment of contagion risk.

\section{Problem Statement and Requirement Analysis}
In this section, we first present the necessary background of this unique financial phenomenon, and then report major concerns obtained from financial experts; finally, we offer a summary of the analysis tasks and design rationales based on real-world motivations.

\subsection{Background of Networked Loans}
\begin{figure}[ht!]
  \centering
  \includegraphics[width=1\linewidth]{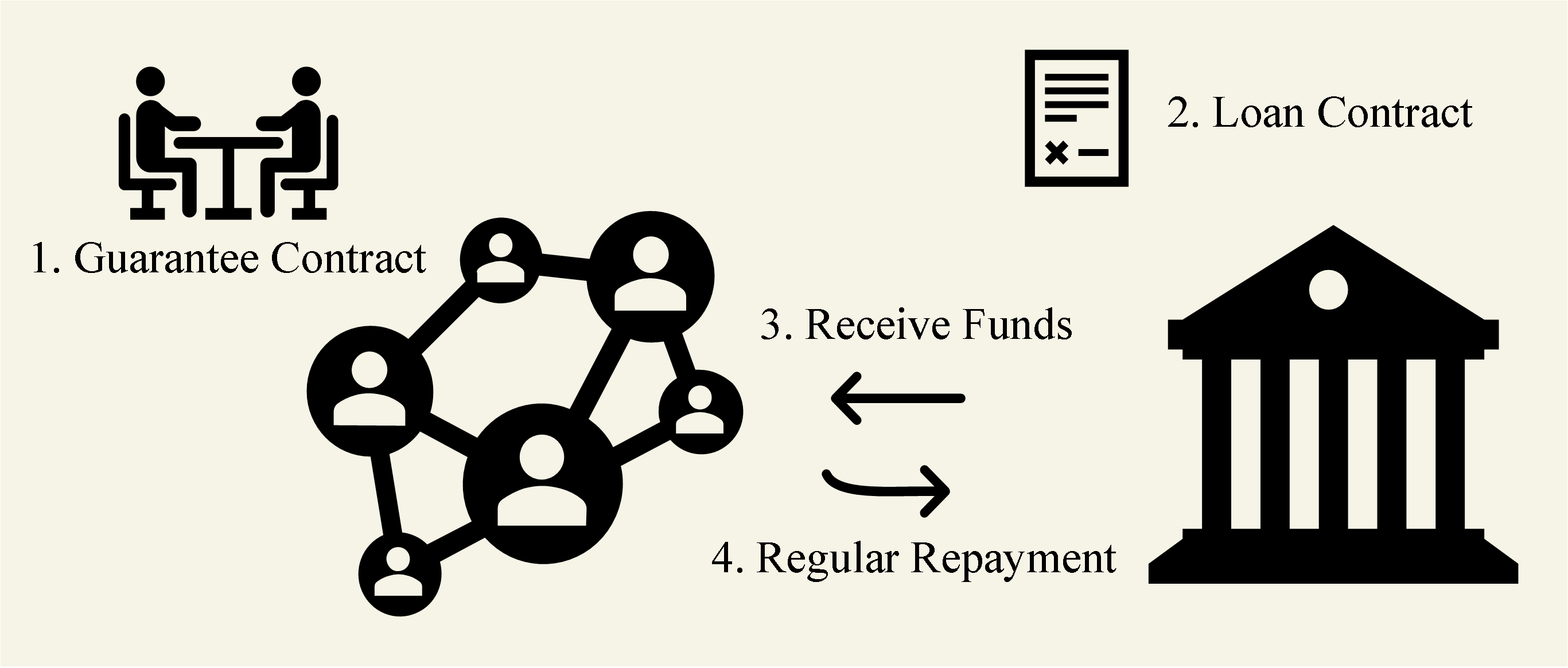}
  \caption{A typical loan guarantee process includes four major steps. 
  }
  \label{process}
\end{figure}

The origin of networked-guarantee loans is illustrated in \autoref{process}. It is a common scenario in which small businesses that wish to obtain loans from commercial banks usually lack the security required. In this case, they are allowed to seek a guarantee from other businesses. In practice, there can be more than one guarantor per loan transaction, and there may be multiple loan transactions for a single guarantor in a given period. Once the loan is approved, the company can usually immediately obtain the full amount of the loan and begin to repay the bank via a regular instalment plan, until the end of the agreement. 


\begin{figure}[ht!]
  \centering
  \includegraphics[width=1\linewidth]{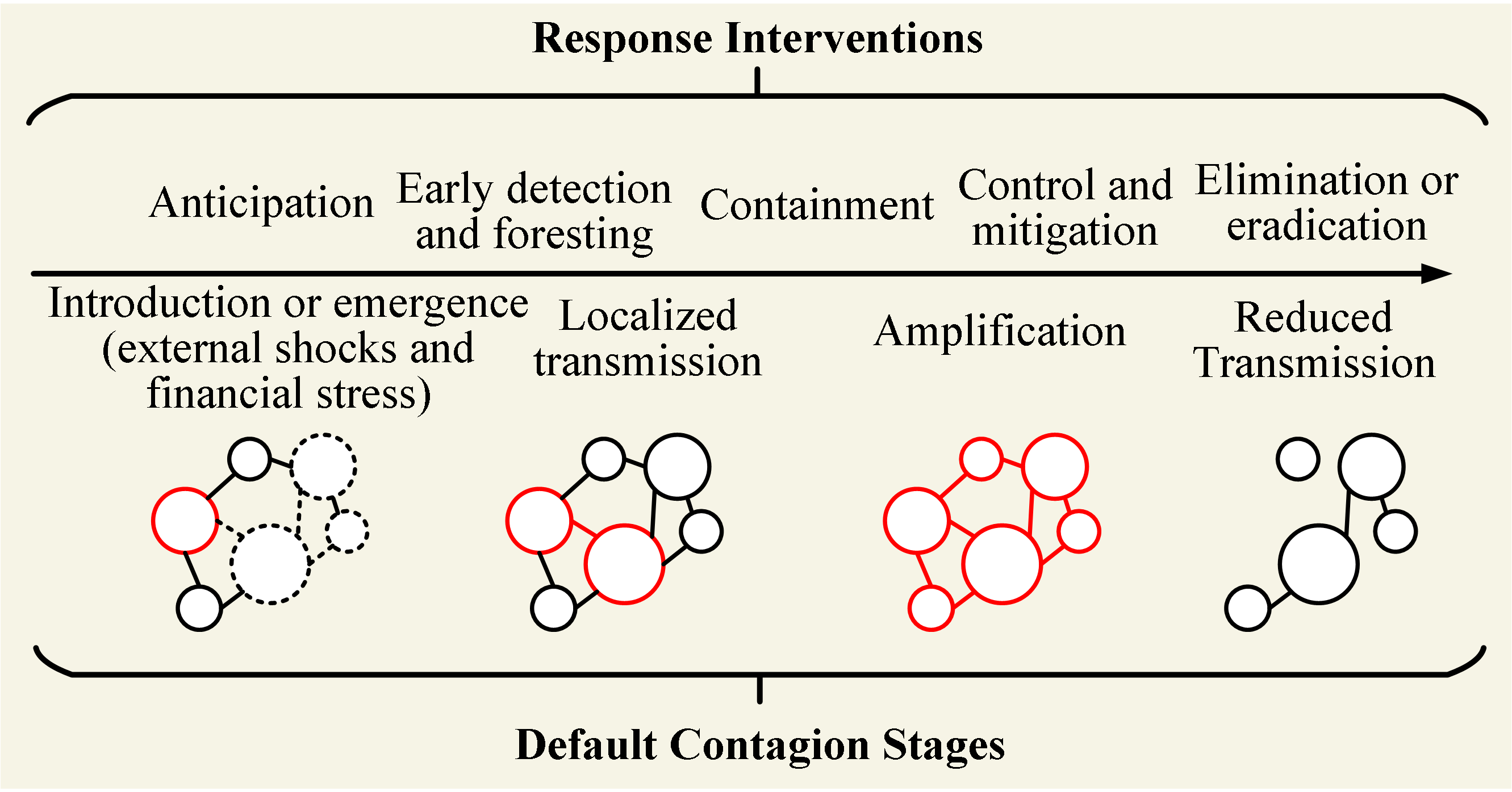}
  \caption{Default contagious phases and response interventions. The red nodes refer to the default company and dashed nodes/links are the default contagion chain.}\label{phase}
\end{figure}

We worked closely with two loan assessment experts to better understand the real-world challenges inherent in this system. Expert $E_{a}$ was a senior financial regulatory officer with more than five years of experience with the guaranteed loan problem and had published several important and relevant investigation reports. Expert $E_{b}$ came from our partner bank, had ten years of loan approval experience. They divided the default contagions and corresponding response interventions into four phases, as shown in \autoref{phase}. 

Default contagions are usually triggered by accidental defaults that introduce risks. Since these are obligation contracts, the default contagion may spread to adjacent nodes (those providing guarantees). \emph{Anticipating how debt default may spread is critical to introducing the appropriate response interventions.} An appropriate guarantee union reduces the risk of default, but in practice, significant damage from contagion can still occur among networked companies. In the case of a down economy, defaults can multiply as large-scale corporate defaults cause side effects in the network. In such cases, the guarantee network can be alienated from the ``joint assistance group'' as a ``breach of contract''. When some companies face operational difficulties, the crisis may set off a domino effect. Default can spread rapidly across the network and put a large number of companies in unfavorable positions. A systemic crisis can result. At this stage, control and mitigation are imperative. After the elimination or eradication stage, the guarantee network may need to be split into several smaller networks, with some companies bankrupted and the transmission risk reduced.

Below, we summarize the main concerns of the experts and outline the requirements for mitigation in the following subsection. \textbf{Concern 1:} Basel accord-based risk management systems may not be well-suited for networked loans, as the network relationship is unique and \emph{exceeds the hypothesis}~\cite{comitato2004international}. It is urgent and essential to adapt the old or establish new risk measurements for this type of problem. \textbf{Concern 2:} Particular attention should be paid to contagion risk to prevent the large-scale corporate defaults often brought about by networked loans. \emph{Accidental default is usually tolerable, while large-scale defaults or systemic financial crises must absolutely be prevented.} However, it is currently unclear how contagion spreads through guarantee networks, how vulnerable the nodes are, or how likely default is to spread. They seek to monitor/assess the status of such spreads via data-driven and visual analytics approaches to resolving risk and ensuring financial stability. 

\subsection{Summary of Analysis Tasks}

Below we summarize our analysis tasks, addressing experts' most significant concerns, supporting the assessment of crisis levels, and gaining insight into precautions preventing potential financial risks.
\vspace{-6pt}

\begin{description} \setlength\itemsep{-0.5em}
\item[T.1] \textbf{Explore networks at different levels of detail, quickly locating networks of interest. } Motivated by expert concerns and technical challenges when analyzing massive bodies of guarantee data, the system should support the quick location of networks of interest and analysis of them according to different levels of detail.
  \item[T.2] \textbf{Understand how default contagion may spread across a guarantee network.}
  The forward approach to understanding the spread of default is to simulate the situation by establishing virus-epidemic models such as in   \cite{pastor2001epidemic}. However, in our case this is impossible, due to a  shortage of empirical default data. From a data-driven perspective, identifying the potential contagion chains and extracting their patterns can also provide insights useful to the ultimate goal of precautions that avoid or resolve systemic financial risk.
\item[T.3] \textbf{Analyze instances of contagion chains. } This should support case-by-case risk assessment and evaluation, as there are multiple instances of contagion chains even with the same pattern. Appropriate quantitative indicators are helpful to making careful comparisons.
  \item[T.4] \textbf{Provide novel and objective risk measurements/indicators tailored for networked loans.} The classic Basel accord-based risk measurement is not well-suited to the problem, as it is based on the assumption that these are giant independent players in the market. It is necessary to set up a novel objective risk indicator for this type of problem.
  \item[T.5] \textbf{Identify the critical nodes that may be the most destructive in case of risk contagion. } Just like the super-spreaders in  epidemiology~\cite{small2006super}, a few corporations are prone to ``superspreading events'', and given the right conditions can ignite explosive epidemics. Moreover, such volatility also means that outbreaks are more likely to fizzle out if the key nodes are identified and removed (as preventive measures) quickly.
\end{description}
\vspace{-6pt}

\begin{figure*}[hbt!]
  \centering
  \includegraphics[width=1\linewidth]{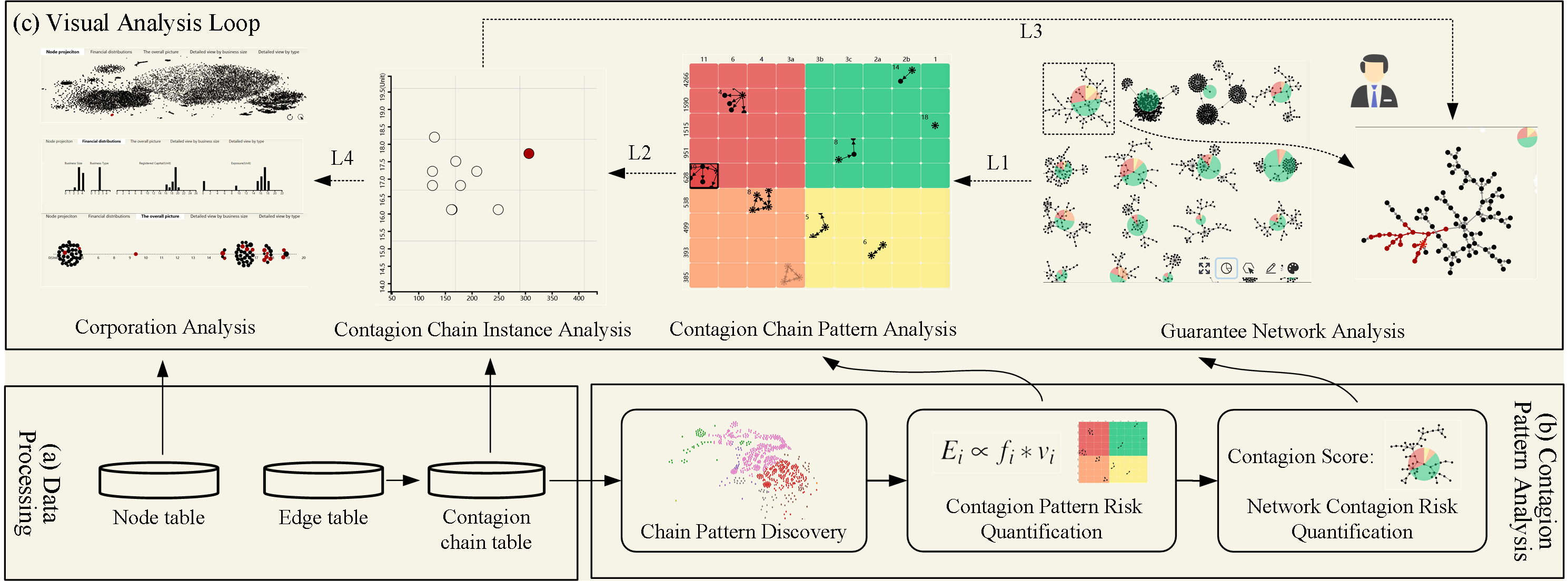}
 \caption{Overview of the approach, including: (a) data preprocessing, (b) contagion pattern analysis, and (c) visual analysis loop. The solid arrow is the data processing flow, and the dotted arrow is the visual analysis flow.}
 \label{method}
\end{figure*}

Based on these analysis tasks, a series of design decisions was made, as outlined below.  \textbf{DR 1. Scalable and appropriate network layout visualization. }The system requires a scalable massive node network layout to avoid severe visual clutter and sufficient performance to enable sophisticated interactions. In practice, target users may wish to analyze networks of interest on various levels of detail; thus, we need a compact arrangement incorporating over 3,000 independent networks via interactions such as magnify, filter, and select (T.1-T.5). \textbf{DR 2. Focus on contagion chains.} Identify and discover   the default contagion patterns in a guarantee network (T.2) and propose appropriate contagion risk indicators/measurements based on the patterns (T.4), as well as intuitive visualization to help with efficient evaluation of the assignment of priorities. \textbf{DR 3. Appropriate symbols and color mapping for intuitive metaphors.} Intuitive representation is an essential element of most visualization systems. Proper visual metaphors help experts reduce the visual burden and improve their understanding of the actual situation (T.1).

\section{Method}
This section describes the method of the visual analytic system.

\subsection{Overview of the Approach}
Motivated by the analytical tasks outlined above, we designed and implemented iConViz to support financial experts interactively exploring the contagion risk associated with networked loans.

\autoref{method} gives an overview of the approach. The process mainly included three steps. 1) Data preprocessing. Raw bank loan records have been cleaned and reorganized into the node table (corporation profiles), edge table (guarantee relations), and contagion chain table. We use solid arrow to illustrate the data processing flow. 2) Contagion pattern analysis. We employed an unsupervised learning-based approach to extract contagion chain structure patterns, and here propose a novel financial metric for quantifying the contagion risk of the chains and networks.  The patterns and metrics formed the basis of our visual design and support this type of financial analysis. 3) Visual analysis loop. We designed a visual interface that is closely coupled with financial risk management tasks and knowledge to support closed-loop analysis processing and an iterative level of detailed exploration (see the dotted arrows and we give more detail in Section 5.5). We describe the details of the data and contagion pattern analysis in the subsequent subsection and the visualization design and interaction in the section after that.


\subsection{Data}
In this work, we collected ten-year loan data from cooperating commercial banks and built the guarantee networks. The names of the customers in the records were encrypted. In the record preprocessing phase, by joining the tables, we obtained records related to the corporation ID and guarantee contract. We then constructed the guarantee network. At the data preprocessing step, we cleaned and reorganized the record data into three main tables, as \autoref{method} shows.

\begin{figure}[hbt!]
  \centering
  \includegraphics[width=1\linewidth]{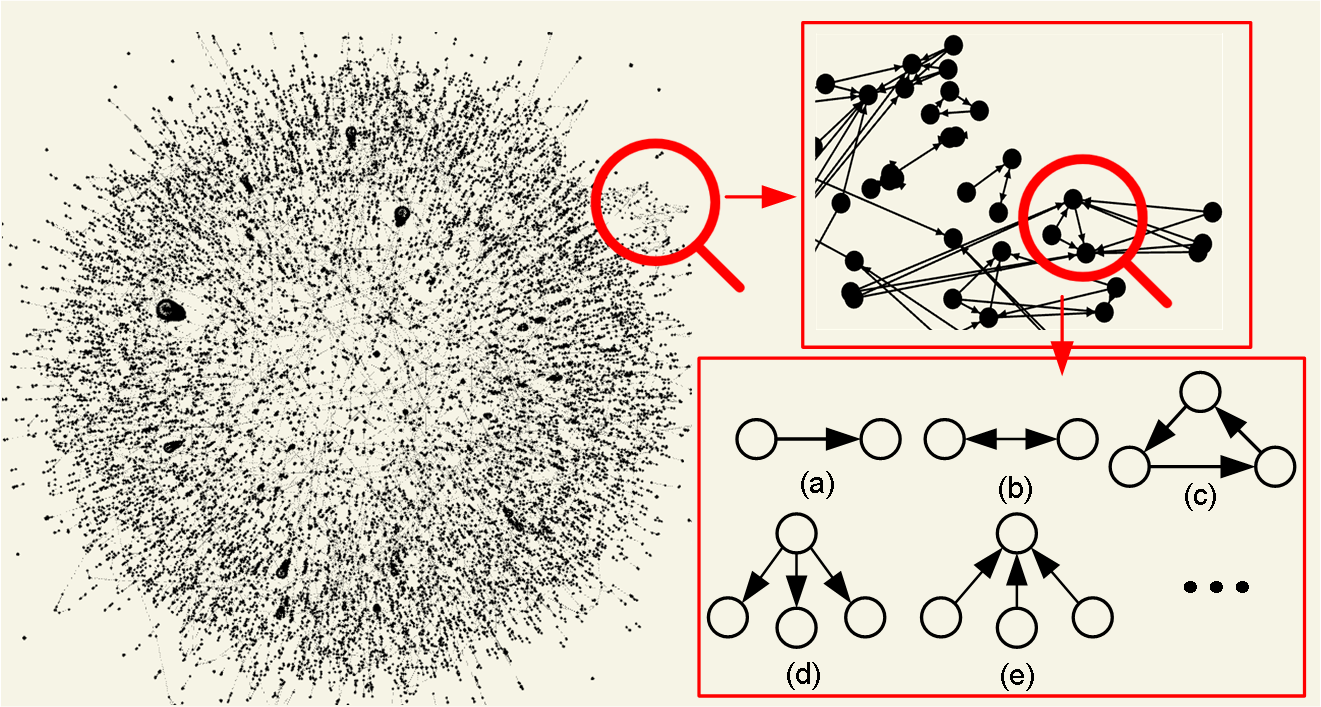}
  \caption{Overview of the real-world dataset. When zooming in, we can observe various directed subgraph microstructure. The classic loan guarantee patterns include~\cite{niu2018visual}: (a) direct guarantee; (b) mutual guarantee; (c) revolving guarantee; (d) star shape guarantee; (e) joint liability guarantee.}\label{hairball}
\end{figure}

1) The node table included the corporation profiles (business type, size, and registered capital); the primary key was the corporation ID. 2) The edge table was the directed guarantee relations between corporations (i.e., the nodes in the node table). It also included the guarantee amounts between them as weights of the edges. \autoref{hairball} gives an overview of the real-world data set, with each node representing an enterprise and the link direction representing the guarantee relationship. More than 20,000 businesses and more than 3,000 independent networks are visualized. It is clear that the networks overlap, and few insights can be drawn. Zooming in, there is a significant directed subgraph microstructure. Some prime aspects may be familiar to loan assessment domain experts~\cite{niu2018visual,cheng2019risk,cheng2019dynamic}, but the collective financial properties of those microstructures are unclear and need to be explored.

\begin{figure}[hbt!]
  \centering
  \includegraphics[width=1\linewidth]{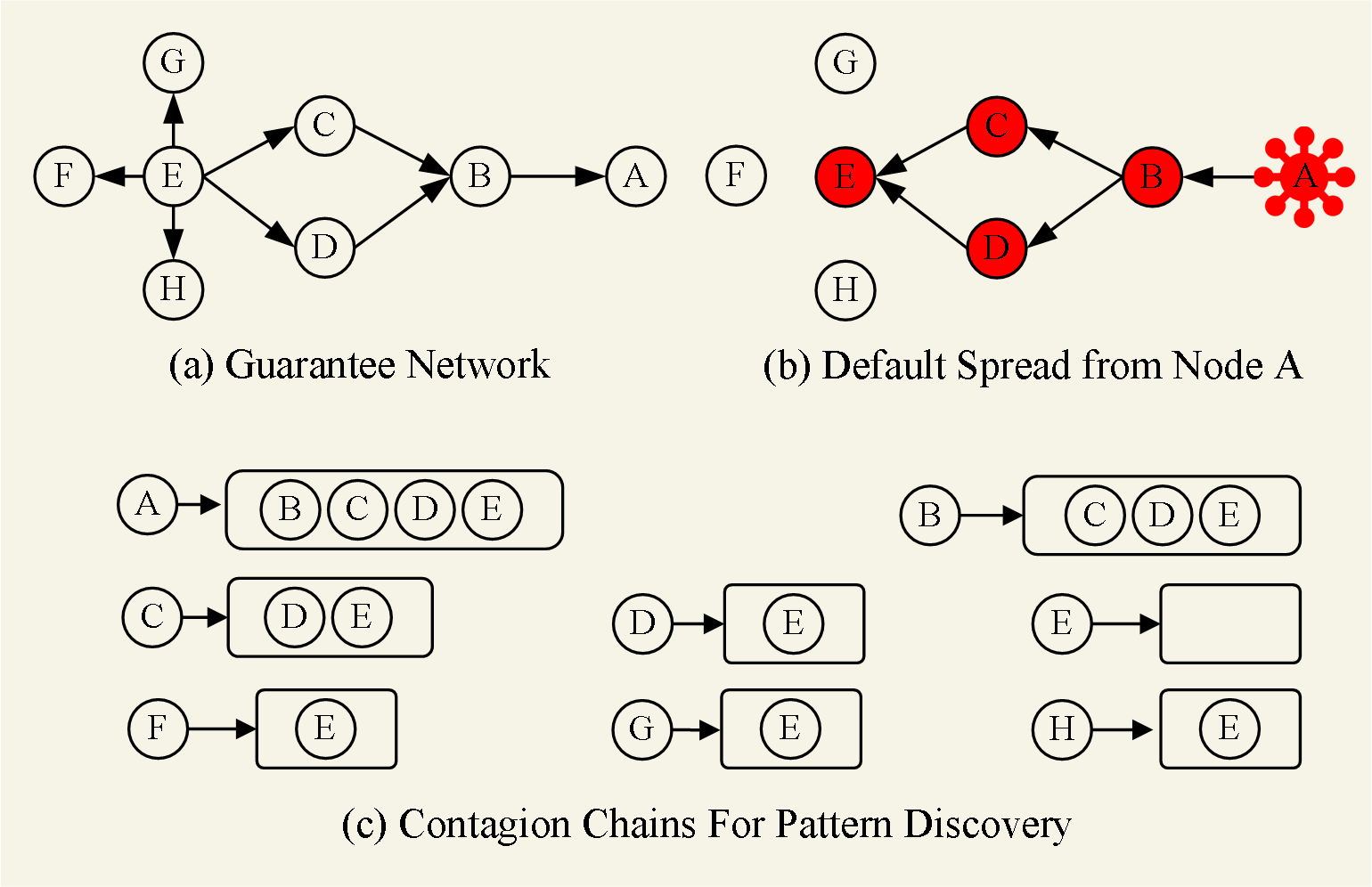}
  \caption{Guarantee network and default contagion chains. In this paper, we utilize the virus-shape node (such as node A in (b)) to represent the source of outbreak (default crisis). }
  \label{contagious}
\end{figure}

3) Contagion chain table. We defined the contagion chain (i.e., the chain of contagion) as the subgraph of where the default might spread. If we can obtain and understand the contagion risk pattern, we may be able to determine how the default risk might spread across the guarantee network and thus implement measures to prevent any potential occurrence of large-scale defaults. The contagion chain table was reconstructed from the edge tables (see the solid arrows in \autoref{method}) and used repeatedly throughout this work. The contagion chain, different from the guarantee chain, worked in the direction opposite to the arrows. In practice, the guarantee network was split into several subgraphs of contagiousness (noted as contagion chains) when we reversed the directions of the arrows. We applied a breadth-first traversal algorithm and generated a series of subgraphs, storing them as contagion chain files in the JSON format. \autoref{contagious} includes an example where (a) is the guarantee network, (b) is the contagion chain when Node A defaults (highlighted with a virus shape icon) and spreads the risk, and (c) gives all possible contagion chains. It should be noted that though some of the chain were subgraphs of other chains, all were analyzed equally because each node could be the source of outbreak. 

\vspace{-6pt}

\subsection{Contagion Chain Pattern Analysis}
\label{s:clustering}

The contagion chain pattern analysis is the basis of our approach and critical to understanding the contagion properties of a network. In this section, we apply the unsupervised learning approach to extract the patterns (typical topological structures), interpret their financial meaning, and quantify the risk brought by each kind of contagion pattern.

The contagion chains were subgraphs, so we extracted network information propagation-related attributes to construct the contagion chain’s features. In this way, each contagion chain was represented by a five-dimensional vector. In detail, the attributes included: (1) the number of nodes and edges of the contagion chain, noted as $N(c_{i})$  and $E(c_{i})$, respectively. (2) The density of the contagion chain, noted as $D(c_{i})$. This was defined as the ratio of the number of edges to the number of possible edges in a network with nodes. It measures the proportion of possible ties that are actualized among the members of a network. (3) The average clustering coefficient of the contagion chain, noted $\overline{C_{i}}$. This is computed as the average of the clustering coefficients $ C_{i} $ of all of the nodes in the chain. It is closely related to the transitivity of a graph and serves as an indicator of a small world. (4) Average shortest path length, noted as $l_G$. This is calculated by finding the shortest paths between all pairs of nodes and taking the average over all paths of the length thereof. It gives the number of steps it takes to get from one node to another and measures the efficiency of information or mass transport on a network.

We chose to use spectral clustering for mining the contagion chain structure patterns. The approach frequently outperforms traditional methods such as k-means or single linkage, especially in graph-based clustering tasks~\cite{von2007tutorial}. Moreover, it can be solved efficiently by standard linear algebra. There are three major steps: Step1: Create the similarity graph between the contagion chains. We chose to fully connect the graph with the Gaussian similarity function to transform a given set $x_{1}, . . ., x_{n}$ of datapoints with pairwise similarities $s_{i,j}$ or pairwise distances $d_{i,j}$ into a graph $G= (V, E)$. Step 2: Compute the first $k$ eigenvectors of the Laplacian matrix to define a feature vector for each contagion chain. The graphed Laplacian matrix is defined as $L = D - W$, where adjacency matrix $W$ is the weight between the vertices of the graph and $D$ is the diagonal degree matrix. It has been proven that the matrix $L$ has as many eigenvalues of zero as there are connected components (or clusters), and the corresponding eigenvectors are the indicator vectors of the connected components. Step 3:Run k-means on these features to separate objects into $k$ classes. Projecting the points into a non-linear embedding enhances the cluster properties in the data so that they can be easily detected. In particular, the simple k-means clustering algorithm has no difficulty detecting the clusters in this new representation with the estimation of $k$ when the eigenvalues are zero.

\begin{figure}[hbt!]
  \centering
  \includegraphics[width=1\linewidth]{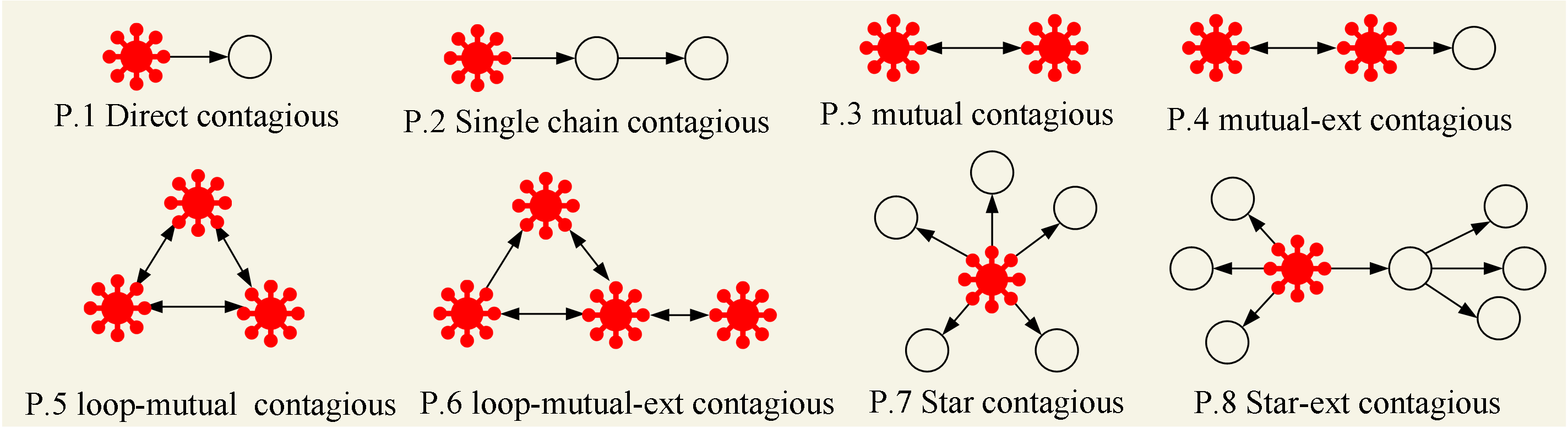}
  \caption{Eight typical contagion chain patterns by unsupervised clustering. The virus shape nodes represent the source of outbreak (the first node in the contagion chain). It is noted that there could be multiple source of outbreaks like in P.3 -- P.6 due to the existence of mutual guarantee.}\label{pattern8}
\end{figure}

\autoref{pattern8} lists the eight basic contagion chain patterns discovered by the above approach. The outbreak (or default crisis) starts from the node in red and spreads through the guarantee network in eight patterns. In detail, they are: \textbf{P.1 Direct contagion pattern.} The basic contagion pattern is where the default can only be spread to its (one) adjacent node and then the contagion is stopped. \textbf{P.2 Single chain contagion pattern.} This usually extends from the direct contagion pattern, with more nodes involved in the contagion chain. The default crisis can only spread across the single chain in the same direction. The length of the entire single strand is arbitrary and all chains in the structure are categorized according to this pattern. \textbf{P.3 Mutual contagion pattern. } This describes a situation in which two corporations simultaneously guarantee one another (mutual guarantees) and obtain funds from a bank. Both nodes are fragile because no matter which one encounters the default crisis, the other will be affected. \textbf{P.4 Mutual-ext contagion pattern.} This is usually an extension of  the mutual contagion pattern, where the contagion chain involves other nodes. \textbf{P.5 Loop-mutual contagion pattern.} This is of a loop structure, when three or more nodes simultaneously guarantee one another. Such a structure is quite vulnerable as any default crisis may spread to all of the nodes in the loop. \textbf{P.6 Loop-mutual-ext contagion pattern.} This is usually extended from the loop-mutual contagion pattern, when the contagion chain involves other nodes. \textbf{P.7 Star contagion pattern.} The default crisis of one node will affect many other nodes and then the contagion will stop. This can occur when corporations provide guarantees for the same (weak) corporation; the default may spread to all of the companies that provide support. \textbf{P.8 Star-ext contagion pattern.}This is extended from the star contagion pattern, but with more complex structures involved. The default of one node may be spread to several other order nodes.

\begin{figure}[ht!]
  \centering
  \includegraphics[width=1\linewidth]{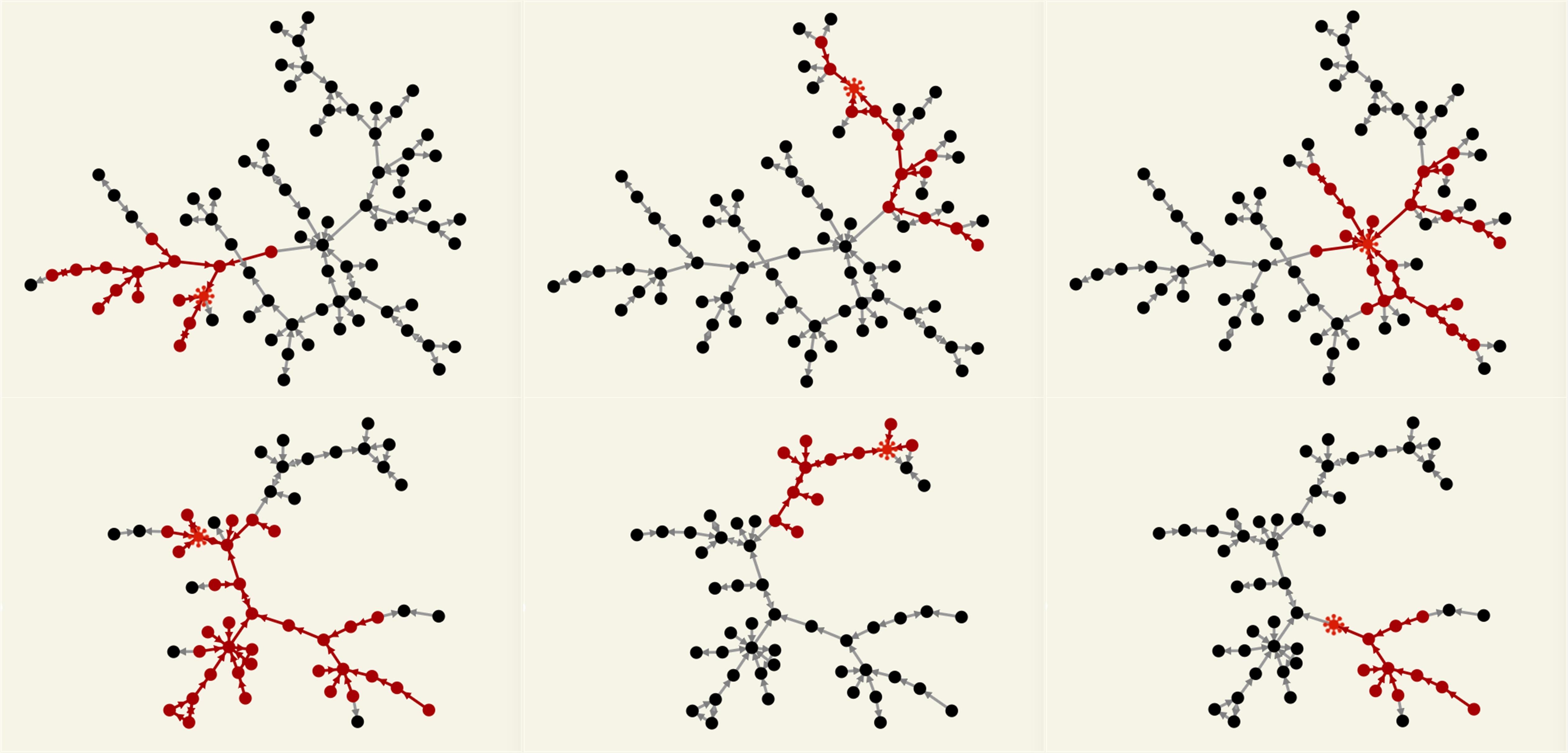}
  \caption{Contagion chains in pattern P.8 in guarantee networks. The clustering algorithm gives each contagion chain a pattern label, and the first nodes of the chains are the source of outbreak. We utilize the virus shape node icon to represent the source of outbreak. }\label{p.8}
\end{figure}

The patterns gradually grow  more complicated with the more nodes involved and more complex the guarantee relationships. In practice, the contagion chains are combinations of these patterns, and a network may have several instances of the same pattern. \autoref{p.8} gives the example of the pattern (P.8) in two guarantee networks. The default spreads across the chain (in red) and then stops. We also observed that these contagion patterns fell into four types, according to the \emph{behavior of the contagion}: single chain, mutual contagious, loop contagious, and star contagious. We propose here a means of quantification for contagion risk assessment based on the behavior of the contagion. Additional details appear in the next section.

\section{Visualization Design and Interaction}
\label{s:interface}
In this section, we describe the interface of the iConViz system that supports financial experts interactively and iteratively exploring and explaining the contagion risk for networked-guarantee loans. We designed the four coordinated views (see \autoref{userinterface}) to facilitate the closed-loop analysis process and iterative level of detailed exploration, following Shneiderman’s mantra. Figure \ref{method} illustrates the analysis procedure. High-level interactions are supported, and all four main views are coordinated to facilitate various levels of detailed exploration of the networks and comparisons of contagion patterns, instances, and specific businesses. It supports a cycle analysis between high-level (massive networks), middle-level (contagion chain-level), and low-level (independent node-level) networks. Such an analysis procedure allows users to finally understand networks and contagion patterns in iterative ways.

\subsection{The Guarantee Network Explorer}
The Guarantee Network Explorer (GNE) view facilities an overview of and zooming in on a level of detail of a guarantee network, using a network tessellation layout. It provides intuitive and metaphorical symbols of contagion risk (through the Contagious Effect Badge (CEB) described in Section 5.2) to support selection by financial interest. It usually works as a starting point. A set of interactive tools are provided to enhance the in-depth analysis of each network.

\emph{Guarantee network tessellation. } Many of the networks are composed of tens or hundreds of nodes, with rare networks composed of thousands of nodes~\cite{niu2018visual}. The naive force-directed graph layout visualizes the whole dataset as a hairball and introduces serious visual clutter. We designed a grid layout to tessellate the guarantee networks, as shown in \autoref{userinterface} (a). In detail, the networks are laid in order of their complexity (i.e., the number of nodes) for the common prejudice that complex networks are more prone to induce large-scale corporate defaults (see Concern 2 in Section 3.1).

\emph{Interactions.} Rich interactions, including brushing, zooming in/out, view panning, and dragging are all supported. The zooming operation supports navigation of the networks at different levels of detail. The panning operation enables the viewing of detailed node profiles. The dragging operation facilitates exploration of the networks in the canvas. Some more enhanced interaction tools have been developed to support in-depth analysis of the networks. At the bottom of the GNE view in \autoref{userinterface}, from left to right, these are: 1) expanded view (\scalerel*{\includegraphics{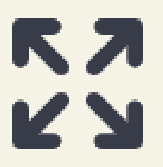}}{\strut}) and 2) risk badge trigger (\scalerel*{\includegraphics{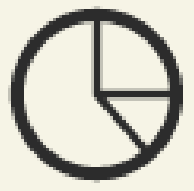}}{\strut}). The risk badges are visual symbols of the contagion risk of the overall network. When the risk badge trigger button is on, all risk badges are overlayed on the networks to help experts locate networks of interest. More details are provided in the following section. 3) Box selection (\scalerel*{\includegraphics{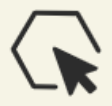}}{\strut}). Users can select the network of interest and highlight the nodes and edges, making the contagion effect matrix (CEM), Chain Instance Explorer (CIE), and Node Instance Explorer (NIE) views display the corresponding content. 4) Edge width trigger (\scalerel*{\includegraphics{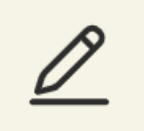}}{\strut}). The edge width trigger is proportional to the guarantee amount. It is useful when analyzing specific networks. We also defined two buttons. The first is to obtain better performance when loading all networks. The second is for use when the width of the edge is not apparent in the network tessellation view. 5) Color palette  (\scalerel*{\includegraphics{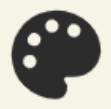}}{\strut}). Nodes can be colored by a default rate (a graph neural network-based prediction that will be discussed in subsequent research), type, and size of businesses.


\subsection{Visual Signaling of Contagion Effect}
Addressing the T.2 and T.4 requirements and based on the behavior of contagion, we created the \emph{Contagion Effect}, a novel financial metric for quantifying the severity of the risk of contagion. Based on this core metric, we designed a Contagion Effect Matrix (CEM) view to \emph{encode the risks} of each network in a matrix manner. It also works as a filter for chain-level analysis. It is further abstracted as the \emph{Contagion Effect Badge}, a visual chart indicating the risk levels appearing in the GNE view.

In finance, the contagion effect explains the impact of a spreading crisis in a situation where one shock in a particular economy or region spreads out and affects other sectors~\cite{aloui2011global,wen2012measuring,baur2012financial,mendoza2010financial,da2016quantifying}. However, as far as we know, there has to date been no measurements quantifying the contagion effect for the guarantee network problem. In this research, we identify the contagion risk associated with networked loans as an important yet unexplored interdisciplinary research problem. As mentioned in the Section 3.1, the classic Basel accord-based risk management systems may not be well-suited for networked loans, as the network relationship is unique and exceeds the hypothesis and accidental default is usually tolerable, while large-scale defaults or systemic financial crises must absolutely be prevented. The extreme case of contagion is the key to preventing any potential systemic crisis when the default spread across a network. Thus, two factors are important: (1). how many corporations a node may affect to the greatest extent possible, and; (2). how frequent different types of contagion occur. With this information, financial experts can locate and prioritize networks of interest. We worked with financial experts to explicitly define the contagion effect of pattern $P_{i}$ as:

\begin{equation}\label{effect}
  E_{i}\propto f_{i}*v_{i}
\end{equation}

where $f_{i}$ is the frequency and $v_{i}$ is the length of the contagion chain pattern. In practice, networks may have contagion chain patterns of various compositions, meaning that default can spread in drastically different ways. The frequency $f_{i}$ describes how much risk is induced by pattern $P_{i}$, and  $v_{i}$ describes how many other nodes it may infect at most. The two key factors ($f_{i}$ and $v_{i}$), though not directly inspired, essentially follow the principle of the risk matrix in the ISO standard risk assessment techniques~\cite{anthony2008s,Riskmatrixiso,Riskmatrixdtu,iso200931010}, where the $Level\ of\ Risk=Probability \times Consequence$.  Quantifying the contagion effect arranges the patterns in the CEM. We used the range of influence (meaning how many other nodes might be impacted) to represent the contagion consequence and level of vulnerability (meaning how frequently this kind of pattern might happen) to represent the contagion probability.

\begin{figure}[ht!]
    \centering
    \includegraphics[width=1\linewidth] {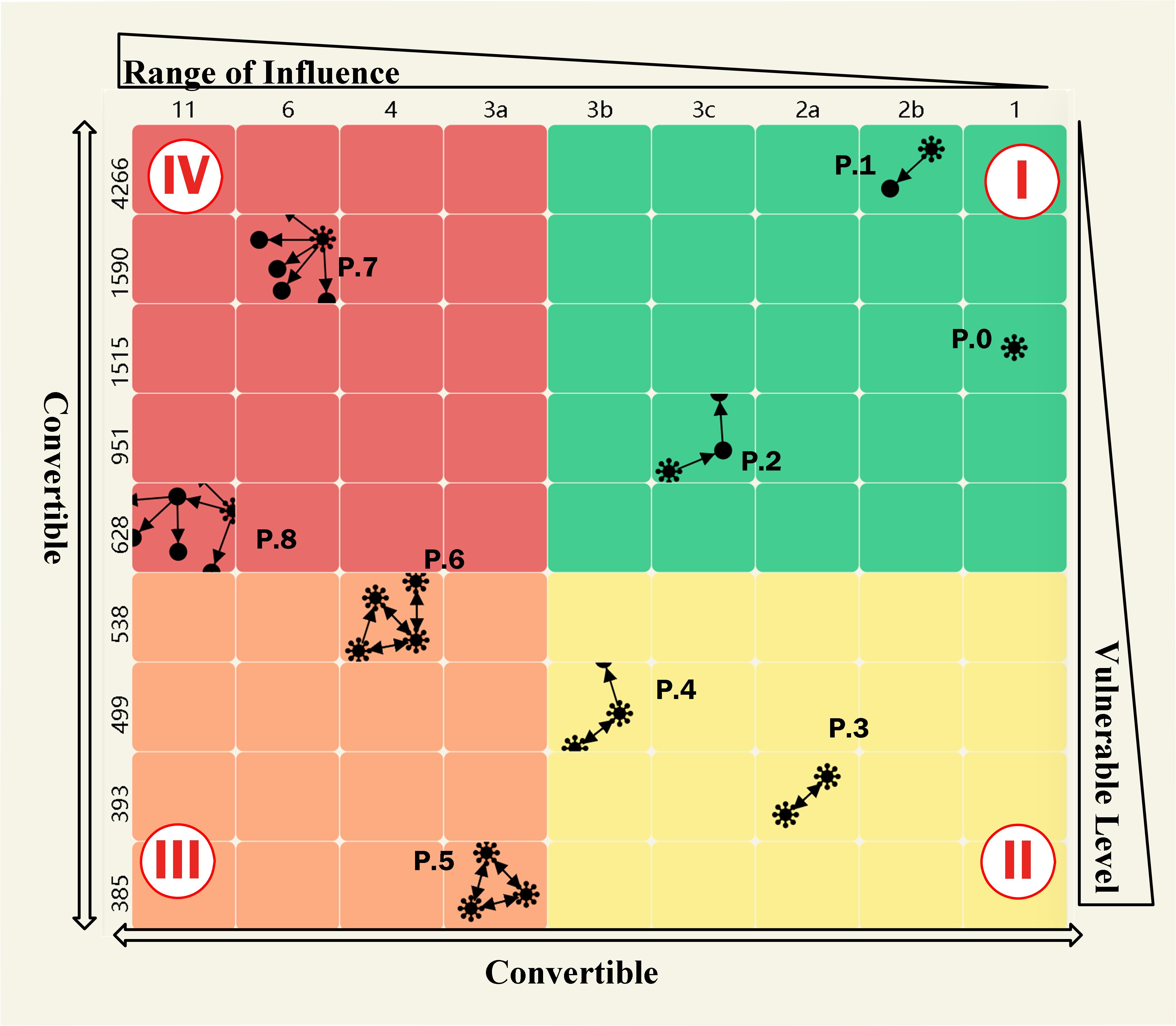}
    \caption{Contagion effect matrix, where the patterns are arranged by the contagion behavior. Each quadrant is given a color according to its contagion risk level. The red, orange, yellow, and green color correspond to high, middle, low, and safe risk levels. This is consistent with the convention settings in the credit rating business. Along the Range of Influence axis is the node number of the patterns. According to the node number, P.1 = P.3, and P.2 = P.4 = P.5, thus, we use the extra letters a, b, and c to distinguish between them.}
    \label{CEM}
\end{figure}

\textbf{Contagion Effect Matrix.} This matrix was designed to visualize the contagion effect of the patterns in a network. As \autoref{CEM} shows, the column is the length $v$ of the contagion chain pattern. The rows are the frequency $f_i$ of the contagion chain pattern, where we directly used the count of instances of the pattern. In this view, all of the patterns are spatially arranged into four quadrants according to the behavior of the contagion. Each quadrant is encoded with a color designating the risk level (consistent with the color specifications in the financial sector). These colors form the basis of the risk badge. Each cell also displays the count of the instances of this pattern for a selected network in the top-left corner.

We categorized these into four kinds of contagion behavior by the range of influence and vulnerability level. Q.I: For chain-like patterns (P.1 and P.2), the default can only spread across a single chain. Usually, such nodes and defaults will not lead to massive defaults. It is relatively easy to break the contagion path by removing the key node on the chain. Q.II: For mutual patterns (P.3 and P.4), the defaults can infect one another (P.3 and P.4). Such patterns are vulnerable because of mutual guarantees. Q.III: For loop-mutual patterns (P.5 and P.6), the default may spread more easily than in chain-like and mutual patterns, due to the existence of loop-mutual guarantees. Q.IV: For star-like patterns (P.7 and P.8), the default in the center chain position may affect all supporting corporations. Such kinds of contagion may distress large numbers of companies.

The layout of the CEM is meaningful. First, the patterns on the left half are more vulnerable to crisis due to the existence of the loop-mutual pattern and broader range of influence, as more nodes are involved in a network than in the right-half patterns. Experts may, in practice, wish to have more guarantee patterns in the right half of the CEM to provide stabler situations. Second, the patterns in the different quadrants can be converted into one another in real situations, providing clues to useful decomposition strategies when splitting a complex network into pieces to avoid potential systemic risks. For example, mutual patterns are more basic than loop-mutual patterns, and when splitting a network during a risk outbreak, we can remove the nodes with a Q.III pattern and generate patterns in Q.II, or even in Q.I.

\textbf{Contagious Effect Badge} Contagion chain patterns pervade each of the guarantee networks. Quantifying the proportional composition can help to identify the type of contagion risk. The patterns can be integrated effectively with financial indicators such as those in the Capital Accord for better risk-level assessments of guarantee networks. We explicitly define the contagion score
as a four-dimensional vector $[EDA,pq_{1},pq_{2},pq_{3},pq_{4}]$, where $EDA$ is the total amount of exposure of the nodes in the network and $pq_{j} $is the percentage share of instances of this kind of contagion behavior.

The CEB is designed as a four-slice pie chart to symbolize the risk levels, based on the contagion score. The size of the CEB is proportional to the relative exposure risk ratio (compared to the maximum exposure of all guarantee networks). The portion of each share of the CEB is encoded as $pq_{j}$ to chart the contagion behaviors. The badge can be overlayed through the functional button onto the networks in the GNE view, allowing users to quickly locate networks of interest (see \autoref{userinterface} (a)). 
The size and color of the CEB encodes the relative exposure ratios and contagion types. Note that the risk levels are not consistent with the complexity of the networks, and users need only choose the network of interest via the CEB. This was emphasized in the training session during the case studies.


\subsection{Chain Instance Explorer}
\begin{figure}[hbt!]
    \centering
    \includegraphics[width=1\linewidth] {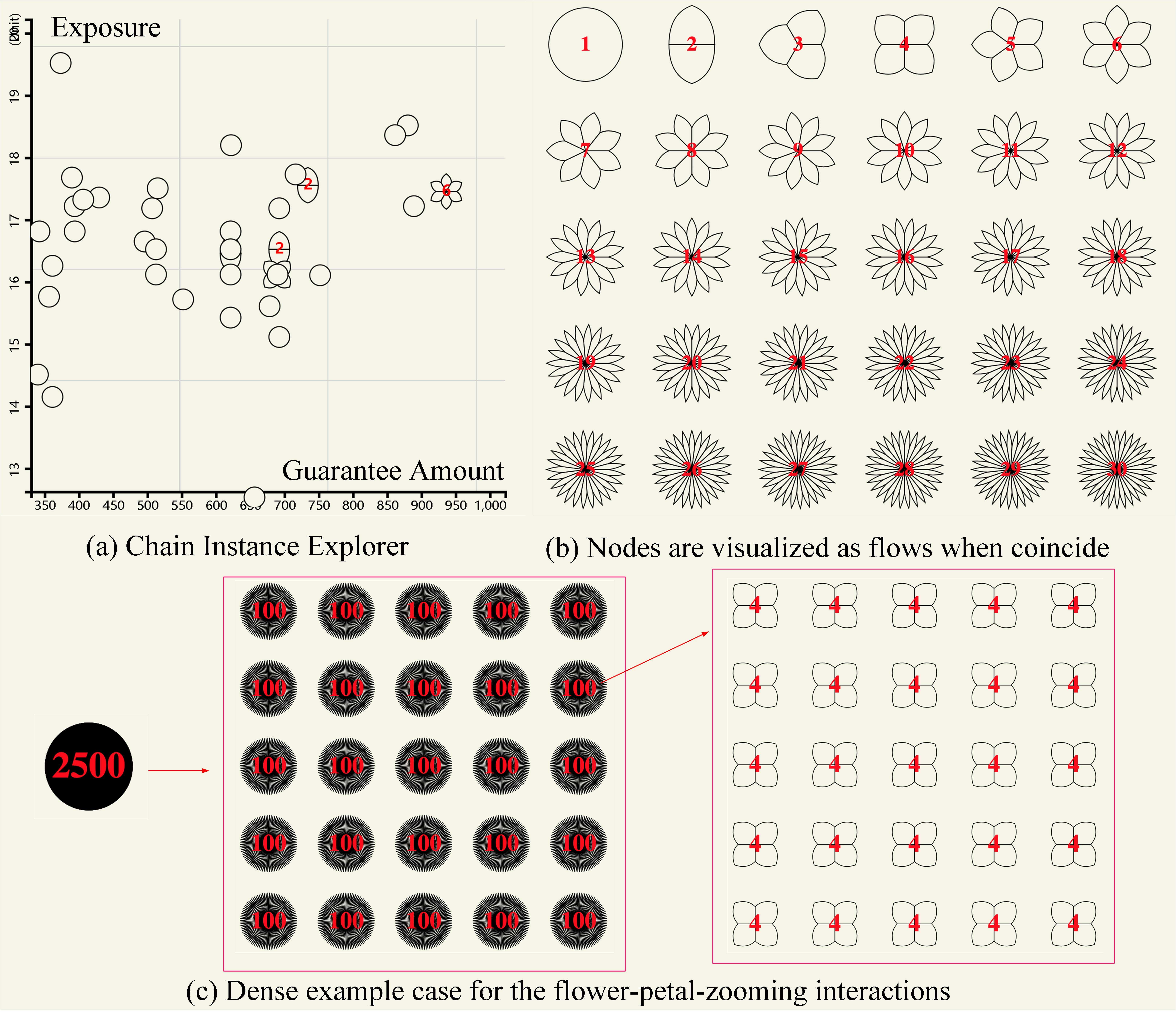}
    \caption{Chain Instance Explorer: (a) financial coordinate system, (b) visualization of contagion chains when multiple nodes are coincident, and (c) dense example case for flower-petal-zooming interactions.}
    \label{CIE}
\end{figure}

The Chain Instance Explorer (CIE) is a tailored financial coordinate system for middle-level (contagion chain) risk analysis. \autoref{CIE} (a) shows the financial coordinate system designed for this research, where the y-axis is the exposure and x-axis is the total guarantee amount of the chain. Each node in the financial coordinate system represents a chain instance. In practice, multiple nodes may have the same exposure and guarantee amount (refer to Section 3.1) and occlude one another. Thus, we propose a \emph{flower-petal-zooming} visualization design for the coincident object selection problem.

In detail, each chain instance is designed as a petal-shaped node in the financial coordinate system. When multiple petals come together, they form a flower. The number in the center shows the count of coincident chains. Each petal is clickable for the user to select the chain instances in other views. The design intension is to avoid possible visual clutter by iterative brush and zooming interactions supported for selection in extremely dense cases (see \autoref{CIE} (c)). The CIE is coordinated with the CEM and GNE, enabling financial experts to explore contagion chains case-by-case.

\subsection{Node Instance Explorer}

\begin{figure}[hbt!]
  \centering
  \includegraphics[width=1\linewidth]{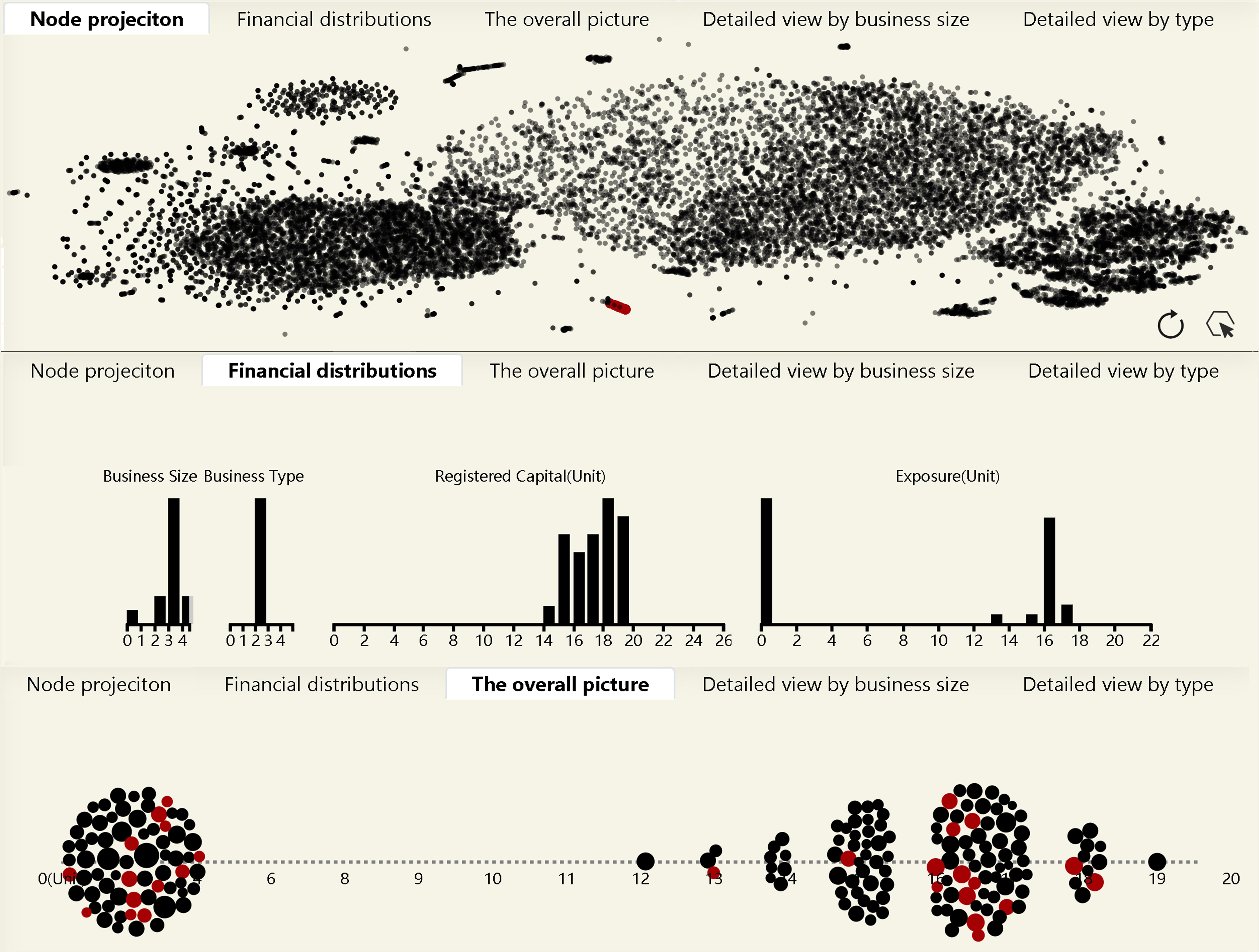}
  \caption{Node Instance Explorer is composed of the (a) node projection view; (b) financial distribution view; (c) overall picture view, consisting of bubbles alongside exposures to give an overview; (d) detailed view by business size and type, which is similar to the overall picture view but omitted due to space limitations.}
  \label{NIE}
\end{figure}

In order to facilitate low-level (i.e., node-level) detail on demand for the finest grain analysis of guaranteed loans, we provide the node instance explorer. It is composed of the following five tabs (see \autoref{NIE}). In detail, they are: 1) Node projection tab.  This provides an overview of the companies by similarities in their guarantee network structures. The nodes are first represented as vectors by node2vec~\cite{grover2016node2vec} and then projected by t-SNE visualization~\cite{maaten2008visualizing}. Box brush interaction is supported for a coordinated analysis. 2) Financial distributions tab. Four important financial statistics (i.e., exposure, registered capital, business size, and business type) are provided. The histograms are visualized by cross-filters to enable a further fine-tuning that includes or excludes records. More sophisticated indicators may be included in the full system when deployed. 3) Overall picture tab. Repulsed bubbles (corporations) are laid along their exposure axis (i.e., their amount of debt) to prevent visual clutter. The bubble sizes are proportional to the corporation’s registered capital. When a user clicks a bubble, the contagion chain is displayed in the GNE. 4) Detailed view by business size and type. Both are charts refined by business size and type to the overview tab.

\vspace{-10pt}
\subsection{Coordinated Interactions}

Coordinated interactions can draw out hidden knowledge from the iConViz system. High-level interactions are supported, and all four main views are coordinated to facilitate various levels of detailed network exploration and comparisons of contagion patterns, instances, and specific businesses. \autoref{method} illustrates the analysis procedure. The system supports a cycle analysis between high-level (massive networks), middle-level (contagion chain-level), and low-level (independent node-level) networks. Such an analysis procedure allows users to eventually understand networks and contagion patterns in iterative ways.

As \autoref{method} shows, the main analysis loop includes four steps. The users start with the GNE view to choose the network of interest with the guidance of the CEB or from the CEM view. They are then able to observe the overall contagion patterns and obtain a ``big picture'' of the data. For example, in the first step from the GNE view, the user can click on the  risk  badge  trigger (\scalerel*{\includegraphics{fig/toolbadge.png}}{\strut}) to overlay the CEB onto the network. A user can then compare the sizes (or radii) of the CEBs to choose the network of interest based on the level of exposure, or compare the proportions of each colored sector to choose the most interesting network based on the contagion behavior. Next, the user can brush-select a network and use the CEM view to see in detail the number of each chain pattern. In the L2 step, a user can click on a cell in the CEM view to see the detailed chain distribution in terms of exposure and total guarantee amount. When the user finds an interesting chain (see the red node in the CIE view in \autoref{method}), they can click on it to see the detailed chain in the GNE view (the L3 step); the source of the break node is marked by a virus-shaped icon (\scalerel*{\includegraphics{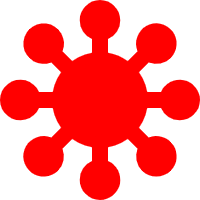}}{\strut}). They can also use the NIE view (the L4 step) to see corporation-level details such as the distribution of financial information.

\section{System Evaluation}


We performed two case studies and one set of expert interviews to evaluate the iConViz system. First, we invited experts $E_{a}$ and $E_{b}$ to conduct case studies to validate the effectiveness of the system in terms of contagion risk assessment. We then conducted a set of in-depth interviews with a broader set of expert users (i.e., risk control managers and specialists) from a cooperative bank.

\subsection{Case Studies}
Experts $E_{a}$ and  $E_{b}$ were organized as a team to perform the case studies. Both had a deep understanding to the real-world financial challenges introduced by guarantee networks. They were also highly involved in our research, as specified in Section 3. They approved the design for the functional view and visual analytics pipeline. Below is a description of their analysis procedure and conclusions.


\subsubsection{Network Level Contagion Risk Assessment}

\begin{figure} [ht!]
  \centering
  \includegraphics[width=1\linewidth]{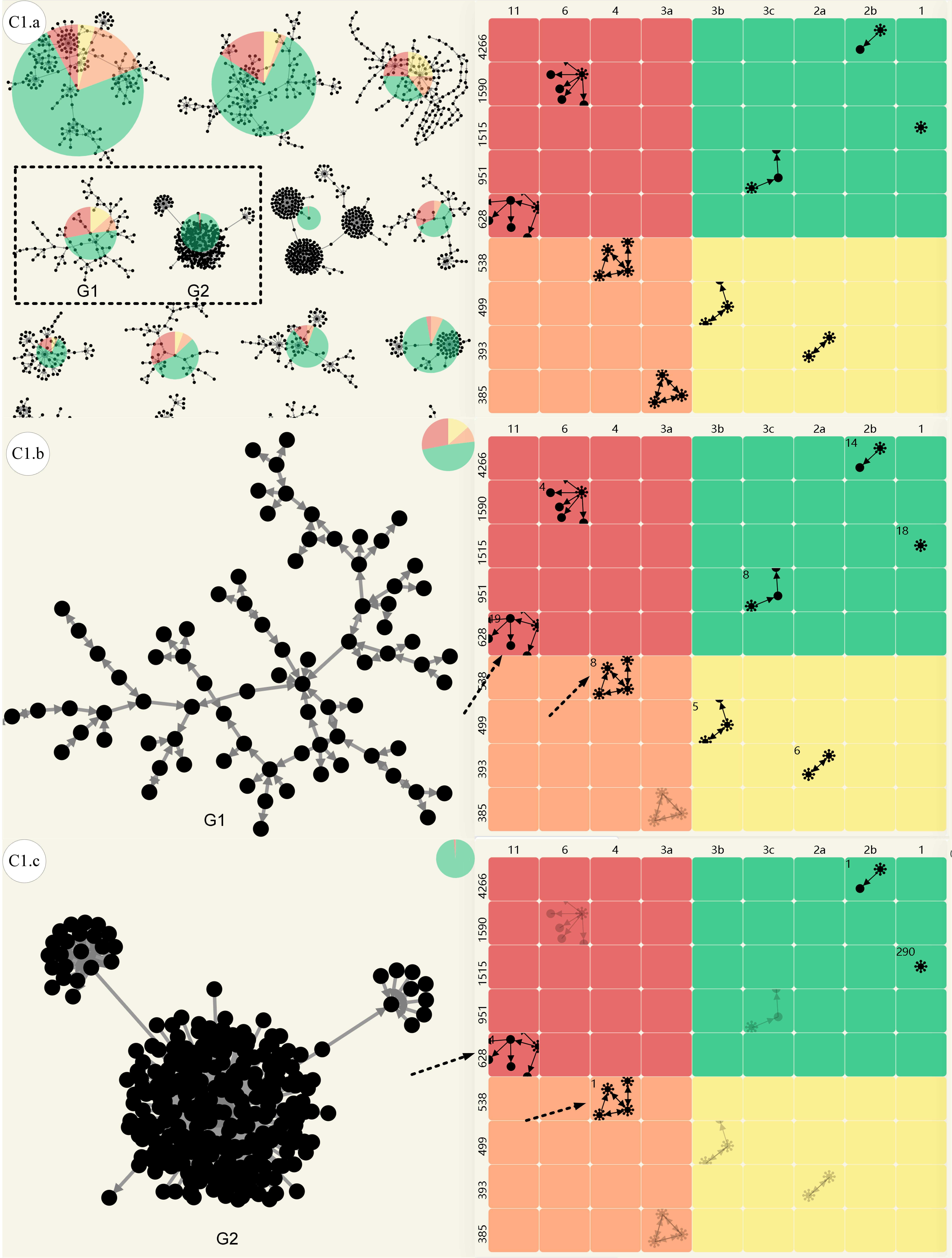}
  \caption{The analysis begins with an overview of contagion risk to facilitate selection and comparison of networks of interest. Zoom in for the numbers.}\label{case_CEB}
\end{figure}

Initially, the experts were attracted to the contagion risk matrix because it was centrally positioned and featured bright colors (see \autoref{case_CEB}, view C.1.a). They familiarized themselves with the red-green credit risk color setting. By default, the numbers alongside the rows indicating the vulnerability levels were arranged by range of influence. They carefully reviewed \emph{the numbers} in the contagion risk matrix (see \autoref{CEM}) and gave overview-level comments on contagion risk (T.2, T.5).

The main insights provided by the experts were as follows.

(a) \emph{The main contagion risk comes from the star-like and loop- mutual patterns in this dataset (T.2).} There was a high number of instances in P.5 to P.8 (as can be observed as 385, 538, 628, and 1,590 on the vulnerability axis), meaning that the revolving and joint liability guarantees (see \autoref{hairball} (c), (e), and Fig.7 for more detail~\cite{niu2018visual}) are common practice, while other patterns such as the star-shaped guarantee (\autoref{hairball}(d)) are relatively stable from the contagion risk perspective. Expert $E_a$ explained that in practice, the star-like guarantee is usually shaped by a professional guarantee company, the joint liability guarantee may be formed by subsidiaries and parent companies, and the revolving guarantee often emerges from business peers of a similar size. Expert $E_b$ remarked that different resolution strategies should be adopted for different kinds of guarantee and contagion patterns. For example, in the P.7 and P.8 contagion instances, in order to prevent risk contagion, the bank should remove the outbreak nodes (see \autoref{pattern8}) so that the network will be broken up into several smaller independent groupings. The P.5 and P.6 contagion instances are a bit more sophisticated, for there might be several potential outbreak nodes. That analysis would need to be coordinated with other iConViz views.

(b) \emph{There are more instances of the P.7 (observed 628 times) and P.8 (observed 1,590 times) patterns than of the P.5 (observed 385 times) and P.6 (observed 538 times) patterns (T.5).} Expert $E_a$ explained that this was caused by the fact that in practice, there are more joint liability guarantees than revolving guarantees. He added that in preventing large-scale corporate default, it is more important for risk control managers to cut contagion chains of the P.7 and P.8 patterns to prevent more businesses being affected.

(c) \emph{The risk is not proportional to the complexity of the guarantee network.} Although the 11 guarantee networks (see \autoref{case_CEB}, view C.1.a) all had complicated network structures, from the size and composition of the badge colors, the experts were able to obtain an overview of how serious the risk was and what types of risk existed in the networks (T.1). For example, the 11 networks had similar node numbers, but the radii of the CEBs obviously varied. This meant that different networks had quite different levels of exposure (EDA for the Contagious Effect badge). Banks should focus first on networks with larger CEB radii, as they have more bank debt (T.4). For instance, the CEB of G1 in C.1.b was obviously larger than that of G2 in C.1.c, meaning that there was more exposure in the former guarantee network and thus it should have attracted more attention from banks. More cases can be observed in \autoref{userinterface} (T.1). Moreover,
the G1 and G2 networks were composed of a similar number of nodes but radically different network structures (see the CEBs). The number in the top-left corner of the CEM (see \autoref{case_CEB}) confirmed that the G1 network in C.1.b had many more star-like patterns ( \scalerel*{\includegraphics{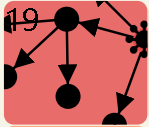}}{\strut} was observed 19 times) and loop-mutual patterns (\scalerel*{\includegraphics{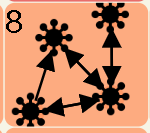}}{\strut}  was observed 8 times). These types can induce a significant risk of contagion during economic downturns (T.2, T.5). Expert $E_a$ affirmed the result after serious consideration, and was pleasantly surprised by the discovery. It makes sense in practice, he explained, because the source of the risk can vary greatly and the classic Basel accord-based risk factors (see Sections 2 and 3.1) cannot accurately describe the risk, especially contagion risk brought about by network relationships. The fundamental issue for risk management is to minimize any potential bank losses. He further affirmed the effectiveness of the contagion score (T4). Expert $E_b$ also confirmed that different from the common perception, networks with more nodes are not necessarily inherently more risky, as network structure and financial information are also critical to such assessments. \emph{Both experts emphasized that this was an important discovery, correcting a misunderstanding in the financial sector that more complex networks are inherently more dangerous.}

The GNE view demonstrates that the CEB can effectively represent risk composition and provide evidence for the selection of analytical priorities (T.1). The conclusions drawn by the financial experts confirmed that the contagion score can provide measurements that perfectly meet the needs of financial professionals (T.4), validating the usefulness of the CEM and extending the visual symbolism of the CEB for quantifying how default contagion may spread across guarantee networks (T.2, T.5).



\subsubsection{Chain Level Contagion Risk Analysis}

\begin{figure} [ht!]
  \centering
  \includegraphics[width=1\linewidth]{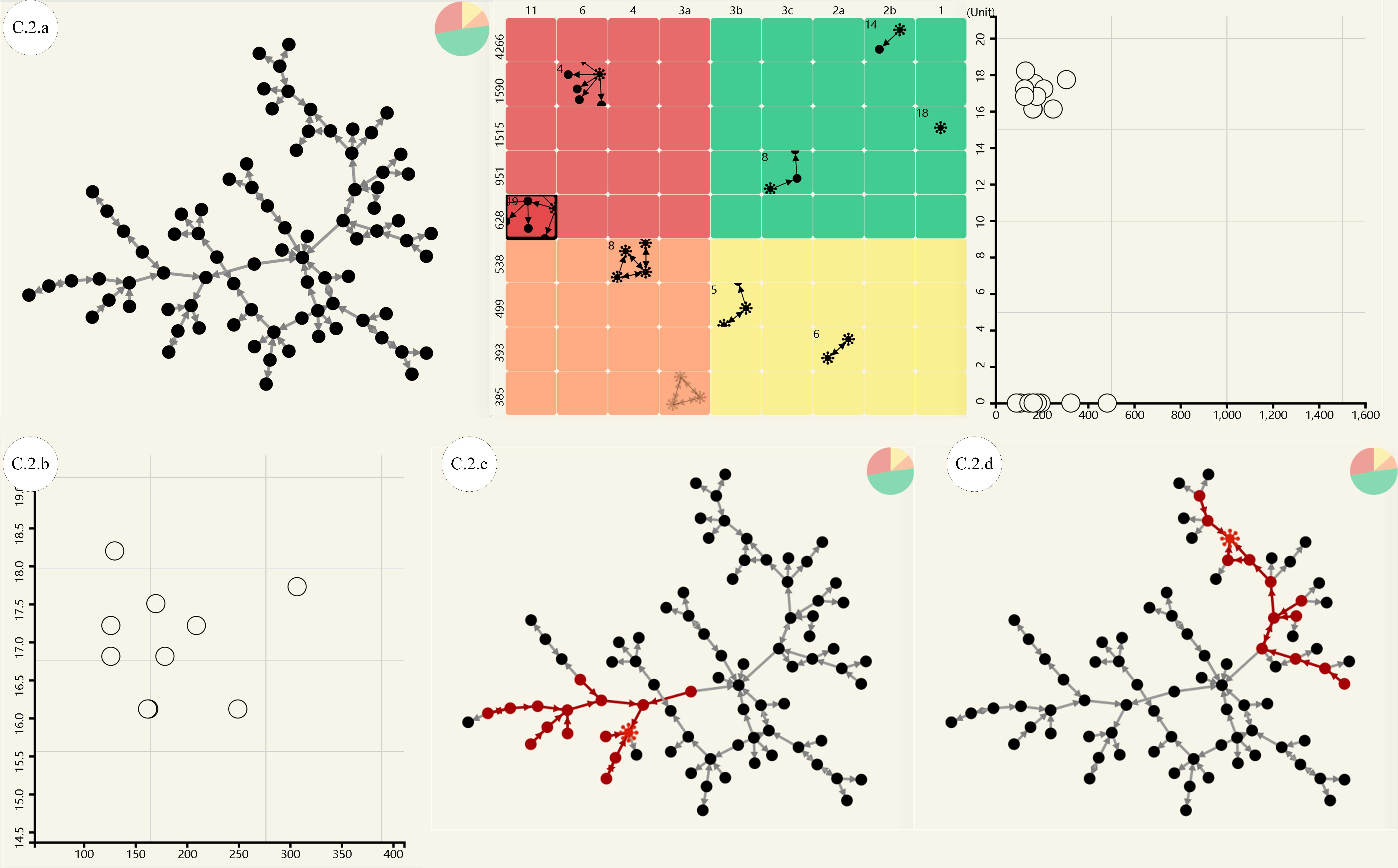}
  \caption{Contagion chain-level risk analysis. The dotted arrow gives the interaction operation. The virus-shaped nodes in views C.2.c and C.2.d were the source nodes for the outbreak. They were automatically labeled by the back-end algorithm.}\label{case1}
\end{figure}

The experts then shifted their attention to chain-level contagion risk (T.1, T.3, T.5). \autoref{case1} shows the interactions and results. The experts brushed and selected the G1 network because it had a relatively large proportion of contagion risk, highlighted in red (\scalerel*{\includegraphics{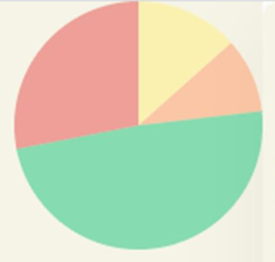}}{\strut}). The network was composed of various contagion chains: 19 instances of the P.8 pattern, eight instances of the P.6  pattern,  and four instances of the P.7 pattern (T.2, T.5). Expert $E_a$ explained that this meant that the network would be vulnerable during a crisis and might influence many nodes. The numbers can thus provide the risk behavior and guide the choice of preventative measures with regards to possible large-scale corporate default. The experts needed to analyze the contagion chains case by case because their financial information was all different and thus they demonstrated different levels of risk. Therefore, the experts chose to analyze the instances in pattern P.8.

As \autoref{case1} shows, in the CIE view, the experts observed that though there were multiple similar chains, some resided on the x-axis, meaning there was no risk exposure (i.e., all the funds were repaid to the bank). There were some nodes in the upper part of the coordinate system that required analysis. Thus, the experts zoomed in on the upper part (the dotted rectangle in the CIE view) and clicked on the nodes to view the chains (view C.2.b). The contagion chains highlighted in red in C.2.c and C.2.d attracted the attention of the experts (T.1), who explained that both were of a complex structure composed of star-ext (see \autoref{pattern8}). Such shapes are multifaceted and difficult to extract via traditional ad hoc examination. Since both types of chain are evidence of vulnerability and have a high range of influence if the default spreads, they need to be addressed, especially during an economic down period. For example, the experts suggested that from the contagion risk perspective, we could retain financial stability by providing individual financial aid or cutting the network at the outbreak node (the virus-shaped node \scalerel*{\includegraphics{fig/virus.png}}{\strut}), as it was in a pivotal position. In practice, preventive measures should be considered in a more sophisticated fashion. The NIE view provides more practical financial information for decision making.

\subsection{Expert Review} 

We conducted a set of expert interviews with target expert users (two risk control managers and five specialists from one bank) to further identify potential problems, validate the design decisions, and assess the system’s effectiveness. All interviewees were familiar with the dataset and had rich experience with loan assessment. We walked them through the system and then let them explore independently throughout the study process. We encouraged them to think about the system and ask questions during the meeting. We interviewed them and collected their comments and feedback regarding the system’s effectiveness, visualization design, and usability in order to identify potential issues and refine the system accordingly.

\textbf{Effectiveness:} The system received high endorsement from all interviewees. Since the risk factors was extended from the widely accepted Capital Accord risk management system, the Risk Badges and Contagion Risk Matrix were easily understood and accepted by the users. The target expert users remarked that it was very useful to be able to quantify the contagion risk and explore the hazards associated with different guarantee networks in such a user-friendly manner. Previously, they had no choice but to adopt a bottom-up approach using Excel or SQL queries (i.e., begin with one node and follow the vine to analyze the neighbors). They could usually only analyze relatively simple guarantee networks and were not provided with the bigger picture. Thus, it was difficult for them to immerse themselves in the massive body of data and analyze their topics of interest. The iConViz system granted them the ability to avoid this analysis strategy. They did not get lost in an ocean of network data. On the contrary, they quickly located their guarantee networks of interest and performed in-depth analyses of the risk patterns. They also expressed interest in deploying the system in the actual risk management procedure for their bank after the study.

We identified several usability issues through the expert user interviews. For example, in our initial version, the system could only visualize the chains in the GNE view when the users clicked on a flower/petal. The chain gave the potential path across which the default might spread. The expert users needed to draw the graph structure on paper and then analyze it. This was inconvenient, so in our updated iConViz system, we automatically labeled the source of the outbreak node using a virus-shaped icon (\scalerel*{\includegraphics{fig/virus.png}}{\strut}) so that the expert users had no problem identifying the key node. The second issue was proposed by a risk control manager who said that these days, deep learning-based risk prediction is popular in bank risk management systems. He suggested that the current iConViz system mainly focus on understanding the existing data and patterns. We will incorporate some deep learning prediction results into our future visual analytics system. We were glad to receive this constructive suggestion and plan to bring prediction results into the iConViz system.

\textbf{Visualization Design:} The experts expressed approval of the overall design. The light-yellow background inspired a sense of spirituality and encouraged creativity. What they most appreciated was the Risk Badge idea. They felt it to be a powerful abstraction of what they wanted to understand. With such intuitive symbols, they did not need to use the mouse wheel to flip through Excel, page by page, and get lost in the data (their words). Moreover, they liked the compact and informative design of the views. They approved of the design of the grid layout for tessellating the guarantee networks, describing it as clear, concise, and providing all of the necessary information. The petal and flower design of the CIE (see \autoref{CIE}) attracted their interest, though during their initial attempts, some ignored this feature. They praised its combined beauty and functionality. However, some users also suggested that it would be more intuitive to give an overview of all the networks on one screen. At that time, they needed to use the mouse to drag the canvas to see all of the networks. There was no slider or other hint in the GNE view. Similarly, we also identified some visualization issues from the feedback. For example, initially, all of the networks in the GNE view had fixed lengths and widths, and it could be difficult to choose those of most interest. We  added a button (expanded view \scalerel*{\includegraphics{fig/toolexpand.png}}{\strut}) so that the user can now choose a full-screen view. Another issue raised by a risk control manager was that he believed the CEM design could be more compact, as it seemed sparse in the middle of the screen. However, he emphasized that the numbers in the CEM were important, as quantitative descriptions are essential in finance.

\textbf{System Interactions:} The expert users believed that the interactions in the views presented by our system were useful and could help them explore and analyze issues important to their work. The interactions among multiple coordinated views facilitated the closed-loop analysis process and iterative level of detailed exploration. Moreover, there were several functional buttons and selection/zooming interactions supported in each view. One expert user said that these were powerful but made the system a bit complex. However, he agreed that it was difficult to make a proper trade-off between complex functions and powerful analytical abilities. For example, without the walkthrough training he would have had no idea how to use the coordinated views to perform the analysis loop, so he suggested that training would be necessary for any future pilot study or deployment. Several risk specialists discovered that due to the sophisticated interactions and massive amount of data being visualized, the system’s operation was not always smooth, and this was its main imperfection. This was because we had to make a trade-off between highly interactive software and the amount of information that could be rendered on a single page. 
Future work will keep improving the interaction performance.

\section{Conclusion}
In this research, we report our progress on risk management in the networked-guarantee loan problem. We believe that this is the first study to identify and formalize the contagion chain risk management problem for networked loans. This new research avenue provides refreshing opportunities for both the computing and financial communities. In our work, a novel financial metric -- the contagion effect, is formulated to quantify the infectious consequences of guarantee chains in a network. Based on this metric, we designed and implemented a series of novel and coordinated views to facilitate analysis of this financial problem. Experts evaluated the system using real-world financial data. We also conducted experts' interview to further collect feedback and improve our system. The results deepen our understanding and ability to assess the potential risk of contagion in complex network structures, hopefully preventing potential large-scale corporate defaults.


\acknowledgments{
This work was supported by National Natural Science Foundation of China (NO.61802278) and foundation of Key Laboratory of Artificial Intelligence, Ministry of Education, China (AI2019004).
}

\newpage


\end{document}